\documentclass[twocolumn,secnumarabic,amssymb, nobibnotes, aps, prl]{revtex4-2}

\usepackage[dvipdfmx]{graphicx}
\usepackage{color}
\usepackage{amsmath}
\newcommand{\red}[1]{\textcolor{black}{#1}}

\begin{document}

\title{Femtosecond reduction of atomic scattering factors triggered by intense x-ray pulse}%

\author{Ichiro Inoue$^{1}$}
\email{inoue@spring8.or.jp}
\author{Jumpei Yamada$^{2}$}
\author{Konrad J. Kapcia$^{3,4}$}
\author{Michal Stransky$^{5, 6}$}
\author{Victor Tkachenko$^{4}$}
\author{Zoltan Jurek$^{4}$}
\author{Takato Inoue$^{7}$}
\author{Taito Osaka$^{1}$}
\author{Yuichi Inubushi$^{1, 8}$}
\author{Atsuki Ito$^2$}
\author{Yuto Tanaka$^{2}$}
\author{Satoshi Matsuyama$^{2,7}$}
\author{Kazuto Yamauchi$^{2,9}$}
\author{Makina Yabashi$^{1,8}$}
\author{Beata Ziaja$^{4,6}$}
\email{beata.ziaja-motyka@cfel.de}

\affiliation{$^1$RIKEN SPring-8 Center, 1-1-1 Kouto, Sayo, Hyogo 679-5148, Japan.\\
$^2$Department of Precision Science and Technology, Graduate School of Engineering, Osaka University, 2-1 Yamada-oka, Suita, Osaka 565-0871, Japan\\
$^3$Institute of Spintronics and Quantum Information, Faculty of Physics, Adam Mickiewicz University in Pozna$\acute{n}$, Uniwersytetu Pozna$\acute{n}$skiego 2, PL-61614 Pozna$\acute{n}$, Poland.\\
$^4$Center of Free-Electron Laser Science CFEL, Deutsches Elektronen-Synchrotron DESY, Notkestr.  85, 22607 Hamburg, Germany.\\
$^5$European XFEL GmbH, Holzkoppel 4, 22869 Schenefeld, Germany.\\
$^6$Institute of Nuclear Physics, Polish Academy of Sciences, Radzikowskiego 152, 31-342 Krakow, Poland.\\
$^7$Department of Materials Physics, Graduate School of Engineering, Nagoya University, Furo-cho, Chikusa, Nagoya, 464-8603, Japan.\\
$^8$Japan Synchrotron Radiation Research Institute, Kouto 1-1-1, Sayo, Hyogo 679-5198, Japan.\\
$^9$Center for Ultra-Precision Science and Technology, Graduate School of Engineering, Osaka University, 2-1 Yamada-oka, Suita, Osaka 565-0871, Japan.}

\begin{abstract}
X-ray diffraction of silicon irradiated with tightly focused femtosecond x-ray pulses (photon energy: 11.5 keV, pulse duration: 6 fs) was measured at various x-ray intensities up to $4.6\times10^{19}$ W/cm$^2$. The measurement reveals that the diffraction intensity is highly suppressed when the x-ray intensity reaches of the order of $10^{19}$ W/cm$^2$. With a dedicated simulation, we confirm the observed reduction of the diffraction intensity is attributed to the femtosecond change in individual atomic scattering factors due to the ultrafast creation of highly ionized atoms through photoionization, Auger decay, and subsequent collisional ionization. We anticipate that this ultrafast reduction of atomic scattering factor will be a basis for new x-ray nonlinear techniques, such as  pulse shortening and contrast variation x-ray scattering.
\end{abstract}
\maketitle

Knowledge of the structure of matter at atomic resolution is critical for understanding and accurately predicting material properties. Since its discovery at the beginning of the twentieth century, x-ray scattering has been a primary tool for atomic-scale structural studies of various systems in physical, chemical, and biological sciences, \red{in particular of crystalline materials} \cite{AlsNielsen2011}.

The recent advent of x-ray free-electron lasers (XFELs) \cite{Saldin1999, McNeilNP2010}, which produce femtosecond hard x-ray pulses, is enhancing the capabilities of x-rays as an atomic-resolution probe.
\red{When an XFEL pulse irradiates solid density matter, atoms undergo sequential electron emission through photoionization and Auger processes, which occur during
the X-ray exposure or shortly after. Subsequently, the ejected electrons interact with bound electrons in neighboring atoms, causing further electron excitations through collisional ionization processes on a timescale of 10 fs or less
 \cite{Ziaja2001, Timneanu2004, Ziaja2005, Caleman2009}.}
Although such electron excitations can trigger  atomic disordering through the electron-lattice interaction and the modifications of interatomic potential \cite{Medvedev2013, Medvedev2015, Medvedev2019}, it has been predicted \cite{Neutze2000} and experimentally confirmed \cite{Inoue2016, Inoue2021, Inoue2022} that there is a several femtosecond time delay  between the x-ray exposure and atomic displacements. Therefore, the ultrafast XFEL pulses allow the measurement of diffraction signal before the onset of the atomic displacements and mitigate radiation damage in the samples,  which has been a long-standing bottleneck for x-ray structure determination \cite{Owen2006, Holton2009, Howells2009, Garman2010}.

Based on this diffraction-before-destruction concept \cite{Chapman2014}, a large number of structures of protein microcrystals have been solved using XFEL pulses \cite{Schlichting2015, Barends2022}. In these experiments, the XFEL pulses were focused to a few micrometer spot size to increase the number of photons irradiating the sample. Even though the intensity and fluence of the microfocused XFEL pulses reached as high as the order of 10$^{17}$ W/cm$^2$ and 10$^{3}$ J/cm$^2$, respectively, no significant electron density gain and loss were observed in the electron density map of the determined structures \cite{Boutet2012}, indicating that the incident photons were predominately scattered by pristine atoms  that had neither been photoionized nor collisionally ionized.

Recent developments in nanofocusing optics for XFEL pulses \cite{David2011, Mimura2014, Seiboth2017, Matsuyama2018, InoueTakato2020, YumotoAS2020} will further strengthen the capabilities of x-ray structure determination. The high intensity and fluence of the nanofocused pulses (more than 10$^{19}$ W/cm$^2$ and 10$^{5}$ J/cm$^2$, respectively) will reduce the required crystal sizes for structure determination and thereby largely expand the targets of x-ray crystallography. However, numerical simulations predict that the majority of the atoms are ionized during the x-ray exposure \cite{Hau-Riege2004, Chapman2014, Medvedev2018} and that the atomic scattering factors become lower than those for neutral atoms \cite{Hau-Riege2007, Son2011}. Thus, it is not appropriate to  use conventional procedures for the structure analysis, and one needs to develop new methodologies that incorporate the ultrafast changes in the atomic scattering factors \cite{Quiney2011}. 

\red{Until now, the response of materials to intense XFEL pulses has mainly been studied using gas-phase atoms and molecules  \cite {Young2018}.
In such systems, the collisional ionization, which is the main electron excitation channel in solid-state materials, is not significant because the majority of the photo- and Auger electrons escape from the system.
Therefore, it is impractical to infer how materials with solid density respond to an intense x-ray pulse based solely on previous studies conducted on gas-phase materials. Although several pioneering groups have applied emission spectroscopy to explore intense x ray-induced electron excitation, particularly focusing on valence electrons in solid and solution samples \cite{Vinko2012, Alonso-Mori2020}, detailed structural studies of x-ray-excited materials with solid density remain unexplored. 
Hence, there are still fundamental questions, such as whether and to what extent the atomic scattering factors are suppressed at high x-ray intensity.}

We describe here an x-ray diffraction measurement of silicon (Si) under irradiation of femtosecond x-ray pulses for different peak intensities and fluences up to $4.6\times10^{19}$ W/cm$^2$ and $3.0\times10^{5}$ J/cm$^2$, respectively. By employing the unique capability of SACLA \cite{IshikawaNP2012} that can generate XFEL pulses with duration well below 10 fs \cite{Inubushi2017, InouePRAB2018, InoueJSR2019},  we measured the x-ray diffraction signals before the manifestation of the x-ray-induced atomic disordering, which becomes prominent at $\sim$20 fs after the x-ray excitation  \cite{Inoue2016, Inoue2021, Inoue2022}, and directly evaluated the change in the atomic scattering factors caused by electron excitations. From the comparison between the experimental results and a dedicated simulation, we discuss the detailed mechanism for the ultrafast reduction of  atomic scattering factors at high x-ray intensity.

\begin{figure}
\includegraphics[width=7cm]{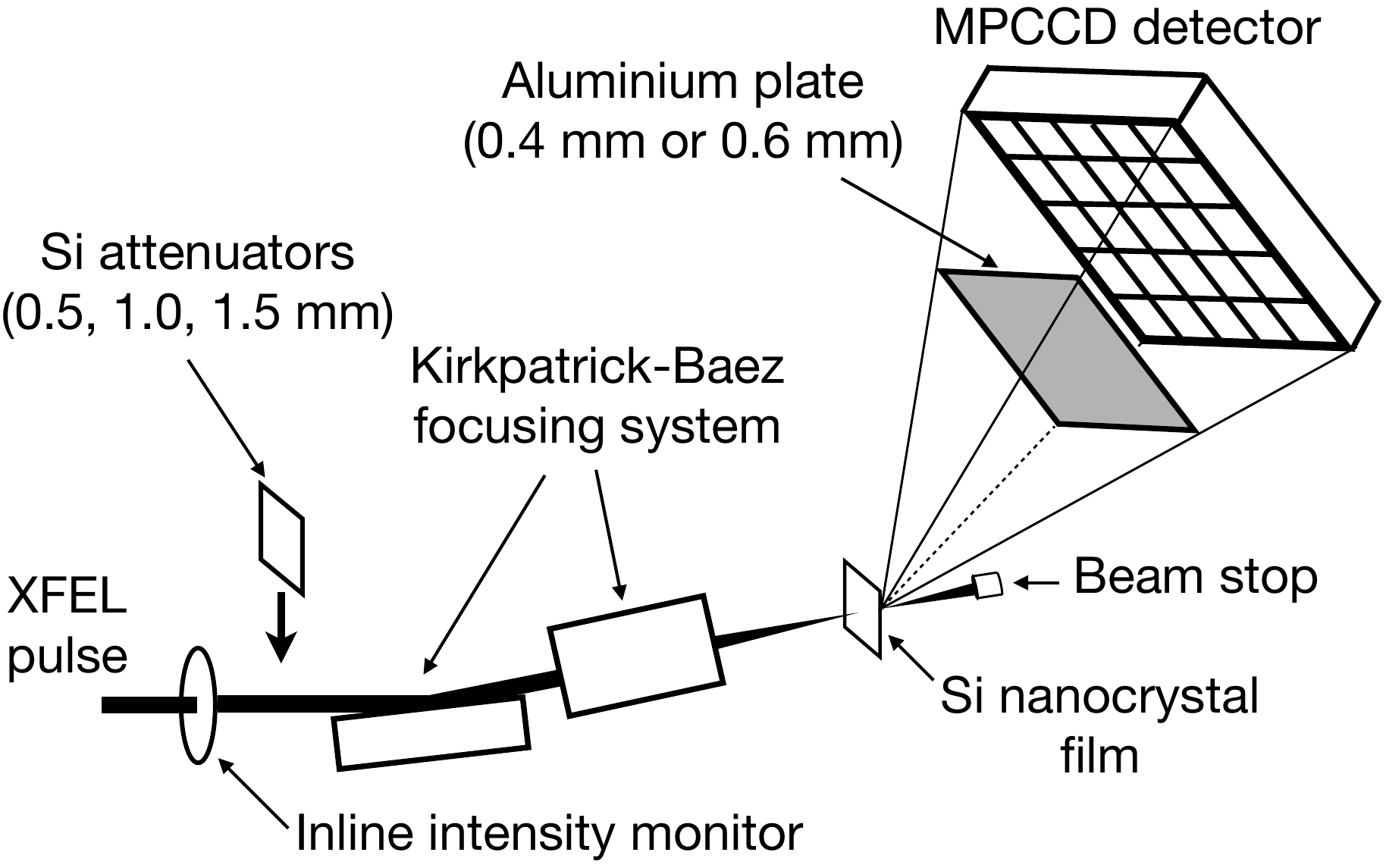}
\caption{A schematic illustration of the experiment.}
\end{figure}

The experiment was performed at experimental hutch 5 of SACLA beamline 3 \cite{YabashiJSR2015, Tono2017} (Fig. 1). The 11.5-keV x-ray pulses with duration of 6 fs were focused by using by using a Kirkpatrick-Baez focusing system \cite{YumotoAS2020}. \red{The full-width-at-half-maximum beam size evaluated by the knife-edge scanning method was 180 nm (horizontal) $\times$ 150 nm (vertical).} A 10-$\mu$m-thick Si nanocrystal film (grain size of 500 nm, US research nanomaterials) attached to a polyimide film was used as a sample. The sample was placed at the focus, and five diffraction peaks (111, 220, 311, 400, 331 reflections) in the vertical plane were measured in a shot-by-shot manner with a multiport charge-coupled device detector (MPCCD) \cite{KameshimaRSI2014} that covered the scattering angle (2$\theta$) range of 18$^\circ$-53$^\circ$. To prevent detector saturation, we placed an aluminum plate (thickness of 0.4 mm or 0.6 mm) in front of the detector. The x-ray intensity at the sample position was tuned by inserting or removing Si attenuators with nominal thicknesses of 0.5, 1.0, and 1.5 mm (measured transmittance was 7.52 \%, 0.60 \%, and 0.045  \%, respectively) before the focusing system. 
The pulse energy at the sample position was monitored by a calibrated inline intensity monitor at the experimental hutch \cite{TonoNJP2013}, taking into account the transmittance of the Si attenuator. The fluence for each pulse was determined by dividing the pulse energy by the product of horizontal and vertical beam sizes (180 nm $\times$ 150 nm). The peak intensity was calculated by dividing the fluence by $\sqrt{\pi/({4\log{2}})} \cdot \Delta t$ with pulse duration $\Delta t$=6 fs.

We compared the diffraction intensities for high and low x-ray intensity conditions as follows. First, we measured 300 successive single-shot diffraction images at the fixed sample position with the 1.5-mm Si  attenuator. The x-ray peak intensity and fluence at the sample were $\sim2.1\times 10^{16}$ W/cm$^2$ and $\sim1.3\times 10^{2}$ J/cm$^2$, respectively.
\red {Given that electron cascade size for an 11.5 keV photoelectron  is $\sim 1$ $\mu$m \cite{Lipp2022_2},  the average x-ray absorbed dose after the electron cascading in measurement with the 1.5-mm Si attenuator was estimated to be 
on the order of 0.01 eV/atom, which is two orders of magnitude smaller than the predicted damage threshold for Si \cite{Medvedev2015, MedvedevPRB2019}.}
From the average diffraction image, we calculated the one-dimensional diffraction intensity profile ($I_{low}(2\theta)$) by azimuthal integration. Next, we reduced the attenuator thickness (0.5 mm or 1.0 mm) or removed the attenuator and measured the single-shot diffraction intensity profile at the same sample position, $I_{high}(2\theta)$. The measurement at low and high x-ray intensities was repeated for different positions.  We extracted and analyzed the dataset of $I_{high}(2\theta)$ measured with specific peak intensity ($(4.6\pm1.2) \times 10^{19}$ W/cm$^2$ (without attenuator),  $(3.5\pm0.9) \times 10^{18}$ W/cm$^2$ (Si 0.5 mm attenuator), and $(2.8\pm0.7) \times 10^{17}$ W/cm$^2$ (Si 1.0 mm attenuator)) and those of $I_{low}(2\theta)$  measured at the same sample positions. For each attenuator condition, the data for $\sim$500 different sample positions were extracted. After being normalized by the pulse energy,  $I_{high}(2\theta)$ and $I_{low}(2\theta)$ were averaged over different positions. Hereafter, we simply refer to these averaged diffraction intensity profiles as diffraction intensity profiles at high and low peak intensities.

\begin{figure}
\includegraphics[width=8.5cm]{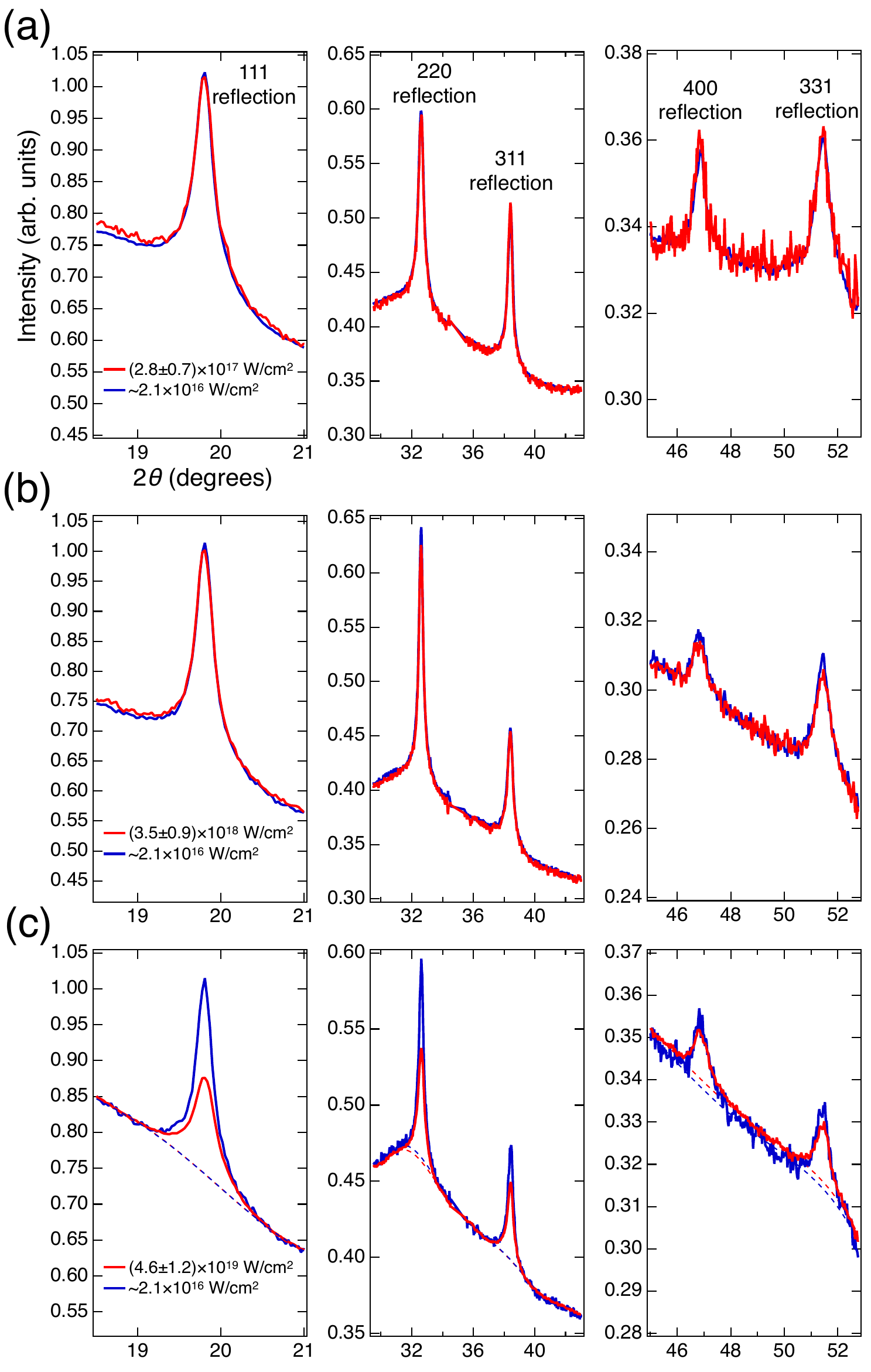}
\caption{X-ray diffraction intensity profiles of Si at high peak intensities: (a) $(2.8\pm0.7) \times 10^{17}$, (b) $(3.5\pm0.9) \times 10^{18}$, (c) $(4.6\pm1.2) \times 10^{19}$ W/cm$^2$, and corresponding diffraction intensity profiles at low peak intensity ($\sim2.1\times10^{16}$ W/cm$^2$). Dotted curves in (c) represent the estimated background.}
\end{figure}

Figures 2 (a)-(c) show diffraction intensity profiles at high peak intensities ($(2.8\pm0.7) \times 10^{17}$,  $(3.5\pm0.9) \times 10^{18}$, $(4.6\pm1.2) \times 10^{19}$ W/cm$^2$) and corresponding diffraction intensity profiles at low peak intensities ($\sim2.1\times10^{16}$ W/cm$^2$). Here we placed a 0.4-mm-thick aluminum plate in front of the detector for the measurement shown in  Fig. 2 (a), while we selected a 0.6-mm-thick plate for  the measurements shown in Figs. 2(b) and 2(c). The background for the diffraction intensity profiles at high and low x-ray intensity conditions was in excellent agreement, indicating that normalization by the pulse energy went well.  As seen from Figs. 2(a) and 2(b), the diffraction intensity profiles at the peak intensities of $(2.8\pm0.7) \times 10^{17}$ W/cm$^2$ and $(3.5\pm0.9) \times 10^{18}$ W/cm$^2$ are almost the same as that at low x-ray intensity. This result proves that the XFEL pulses with the intensities of up to $10^{18}$ W/cm$^2$ do not change the atomic scattering factors and the degree of atomic disordering during the x-ray exposure, validating damage-free protein crystallography using microfocused XFEL pulses (typical intensity and fluence are 10$^{17}$ W/cm$^2$ and 10$^3$ J/cm$^2$, respectively), which is routinely performed at XFEL facilities \cite{Schlichting2015, Barends2022}.

In contrast, the diffraction intensity at the highest peak intensity ($(4.6\pm1.2) \times 10^{19}$ W/cm$^2$) was suppressed compared with that at low peak intensity (Fig. 2(c)). The observed decrease in the diffraction intensity indicates structural and/or electronic damage in Si crystals during the x-ray exposure. To quantitatively evaluate how much the diffraction intensity was suppressed at high x-ray intensity, we first estimated the background of the diffraction profiles at high and low x-ray intensity by fitting the profiles in the vicinity of diffraction peaks with polynomial functions (dotted curves in Fig. 2(c)). After subtracting the estimated background, each diffraction peak was fitted by a Gaussian function, and the integrated diffraction intensity for $hkl$ reflection ($hkl$=111, 220, 311, 400, 331) was determined ($I_{high}^{hkl}$ and $I_{low}^{hkl}$). Table 1 summarizes the ratio of the diffraction intensity at high intensity to that at low intensity $I_{eff}^{hkl}=I_{high}^{hkl}/I_{low}^{hkl}$ (hereafter called the diffraction efficiency) and corresponding scattering vector $Q=4\pi \sin \theta/\lambda$ with the x-ray wavelength $\lambda$. The experimental uncertainty of $I_{eff}^{hkl}$ in Table 1 represents the standard deviation of the diffraction efficiency calculated for five independent sub-ensemble datasets.  The diffraction efficiency did not depend much on $Q$, indicating that the atomic disordering during the pulse irradiation (which reduces the diffraction intensity more at higher $Q$ values \cite{Barty2012}) was not significant in the present experiment. Thus, it is natural to consider that the observed reduction of the diffraction intensity was attributed to the femtosecond change in atomic scattering factors due to the electronic excitations induced by the XFEL pulse.

\begin{table}
\caption{
Ratio of x-ray diffraction intensity of Si normalized by incident pulse energy at high peak intensity ($(4.6\pm1.2) \times 10^{19}$ W/cm$^2$) to that at low peak intensity  ($\sim2.1\times10^{16}$ W/cm$^2$).}
\begin{ruledtabular}
\begin{tabular}{cccccccc}
Reflection ($hkl$)&$I_{eff}^{hkl}=I_{high}^{hkl}/I_{low}^{hkl}$ & $Q$ (\AA$^{-1}$)
\\
\hline
111 & 0.650$\pm$ 0.040 & 2.00 \\
220 & 0.735$\pm$ 0.084 & 3.27 \\
311 & 0.760$\pm$ 0.138 & 3.84 \\
400 & 0.769$\pm$ 0.169 & 4.64 \\
331 & 0.724$\pm$ 0.046 & 5.04 \\
\end{tabular}
\end{ruledtabular}
\end{table}

To justify this statement, we performed simulations of Si crystal under irradiation with a 6-fs XFEL pulse, using the released version of the molecular dynamics code, \textit{XMDYN}  \cite{Jurek2016,Son2011,Murphy2014}. Neutral atoms, atomic ions, and ionized electrons were treated there as classical particles, and their real-space dynamics were calculated by molecular dynamics technique. The electronic configurations of atoms and ions were followed by taking into account all relevant x-ray-induced processes in matter (such as photoionization, Auger processes, fluorescent decay, and collisional ionization and recombination). Although the focal spot of the XFEL pulses had a Gaussian shape in the present experiment, and the diffraction signals originated from various sample areas with different fluence,  we performed the simulations assuming uniform x-ray fluence to reduce the computational cost.

The simulation was performed for the incident pulse with a constant pulse duration (6 fs) and different peak intensities.
\red{After normalization of the simulated diffraction intensity by the incident pulse energy,  $I_{eff}^{hkl}$ was calculated by dividing the normalized diffraction intensity by that for peak intensity of $2.1 \times 10^{16}$ W/cm$^2$.
We found that the simulation results for the peak intensity of $1.0 \times 10^{20}$ W/cm$^2$ could well reproduce the trend of $I_{eff}^{hkl}$ shown in Table 1 (Fig. 3(a)).}
In Fig. 3(a), the error bars for the simulation results represent the deviation of maximum and minimum values of $I_{eff}^{hkl}$ from the average value obtained from ten independent \textit{XMDYN} simulations. Similarly to the experimental observations, the simulation predicts a nearly constant decrease in the diffraction efficiency for the five reflections ($hkl=$111, 220, 311, 400, 331). Figure 3(b) shows the simulated root-mean-square atomic displacement during irradiation with the XFEL pulse at the same intensity. For reference, the temporal intensity envelope of the XFEL pulse is also shown. The atomic displacement is much less than lattice spacing for the measured reflections, indicating that the reduction of the diffraction efficiency is not caused by x-ray-induced atomic disordering. Figs. 3(c) and 3(d) show the relative ion population of Si atoms and the average hole numbers in $K$, $L$ and $M$-shells per atom. It is clearly seen from Fig. 3(c) that the majority of Si atoms are no more neutral during the x-ray exposure. Furthermore, the x-ray-induced electron excitations are not limited to valence electrons  (Fig. 3(d));  many atoms have vacancies in their inner-shells ($K$ and $L$-shells). 
\red{Since the atomic scattering factor is related to electron density distribution via Fourier transformation \cite{AlsNielsen2011}, the x-ray scattering by valence electrons occurs only in the forward direction with small scattering angles.
The atomic scattering factors of neutral and ionized Si atoms at scattering vector corresponding to Bragg reflections are predominantly determined by the electronic occupation of atomic inner-shells. 
Thus, the massive excitation of inner-shell electrons reduces the diffraction efficiency.}
The simulation results support our hypothesis that the suppression of the diffraction intensity at high x-ray intensity of the order of 10$^{19}$ W/cm$^2$, which was observed in the experiment, was caused by the reduction of atomic scattering factors triggered by x-ray-induced electron excitations. 
\begin{figure}
\includegraphics[width=8.5cm]{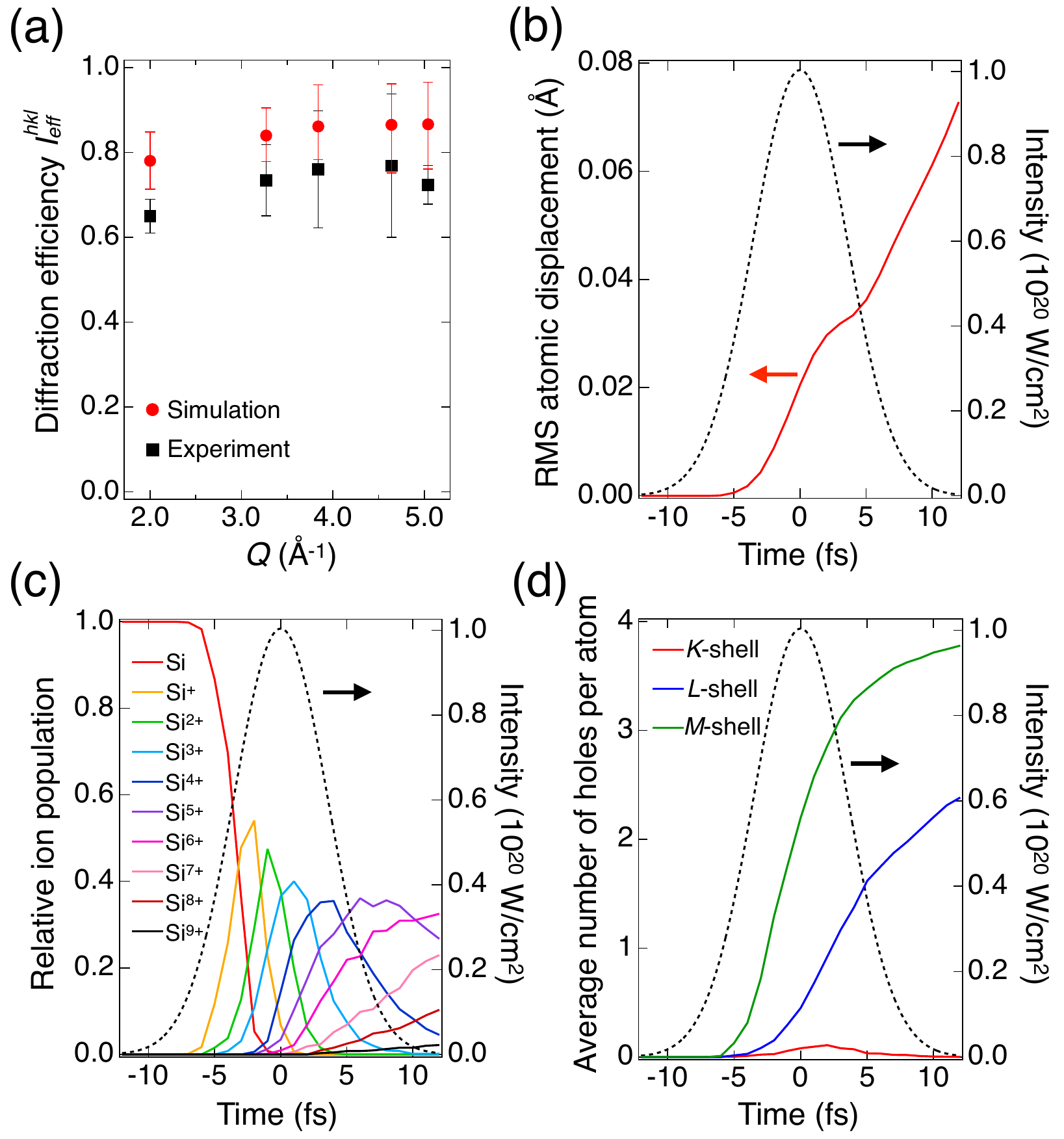}
\caption{\textit{XMDYN} simulation results for Si crystal during its exposure to 6-fs XFEL pulse with photon energy of 11.5 keV and peak intensity of $1.0\times 10^{20}$ W/cm$^2$. (a) Comparison of diffraction efficiency for 111, 220, 311, 400, 311 reflections of Si obtained from simulations (red circles) and from experiment (black squares). (b-d) (b) Root-mean-square atomic displacement, (c) relative ion population, and (d) average numbers of $K$,$L$, and $M$-holes per atom, for Si exposed to the XFEL pulse. Time zero corresponds to the intensity maximum of the XFEL pulse. Black dotted curves represent temporal intensity envelope of the XFEL pulse.}
\end{figure}

In summary, we measured the x-ray diffraction intensity of Si under irradiation of nanofocused 11.5-keV XFEL pulses. The measurement reveals that diffraction intensity is suppressed at the  x-ray intensity on the order of $10^{19}$ W/cm$^2$. From a nearly constant decrease in the diffraction efficiency for the five reflections and a dedicated simulation, we concluded that the reduction of diffraction intensity is attributed to femtosecond change in individual atomic scattering factors due to x-ray-induced electron excitations. 
We anticipate that the ultrafast reduction of the atomic scattering factors can be a basis for novel applications of high-intensity XFEL pulses. 
One intriguing application is the nonlinear optical device for pulse shortening in the hard x-ray regime \cite{Inoue2021_2}. The simulations results show in Fig. 3 predict that Si crystals under high-intensity x-ray irradiation become highly ionized, and thereby its overall scattering strength is largely suppressed in the second half of the pulse. Therefore, the photons at the leading edge of the pulse are selectively diffracted, effectively making the pulse duration of the diffracted beam shorter than the incident pulse duration. Since the suppression of atomic scattering factors at high x-ray intensity is expected to occur in all materials, thin single crystals made of light elements in Laue geometry are promising candidates for high-throughput optical devices designed to shorten the pulse duration. \red {Such devices will enable production of Fourier-transform-limited x-ray pulses}. Another potential application is the contrast variation x-ray scattering in polyatomic samples. Since the electron-impact ionization cross-sections depend on atomic number \cite{Lotz1968}, we can expect that the magnitude of decrease in the atomic scattering factor at high x-ray intensity will differ from atom to atom. The diffraction measurement at different x-ray intensities will then enable contrast variation of scattering strengths between different types of atoms \cite{Son2011PRL, Son2013}. This can open a new route for  \textit{de novo} structure determination of protein crystals.

 \ \ 
\acknowledgements{We acknowledge Prof. Eiji Nishibori and Dr. Kenji Tamasaku for  insightful discussions. The work was supported by the Japan Society for the Promotion of Science (JSPS) KAKENHI Grants (19K20604, 22H03877). K. J. K. thanks the Polish National Agency for Academic Exchange for funding in the frame of the Bekker programme (PPN/BEK/2020/1/00184). The experiments were performed with the approval of the Japan Synchrotron Radiation Research Institute (JASRI, Proposal Nos. 2021A8057, 2021B8022, 2022A8030).}

\bibstyle{natbib}
\bibliography{Ref}

\end{document}